\documentclass[conference, 10pt]{IEEEtran}

\usepackage{cite}
\usepackage{url}
\usepackage{graphicx}
\usepackage{color}
\usepackage{placeins}
\usepackage{float}
\usepackage{tabularx,colortbl}
\usepackage{ifthen}
\usepackage{tikz}
\usepackage{amsmath,amsfonts,amsthm,amssymb,esint} 
\usepackage{subcaption}
\usepackage[
top    = 21mm,
bottom = 1.3in,
left   = 0.62in,
right  = 0.62in]{geometry}
\definecolor{darkblue}{rgb}{0.0,0.0,0.3}
\definecolor{darkred}{rgb}{0.3,0.0,0.0}
\definecolor{darkgreen}{rgb}{0.0,0.3,0.0}

\makeatletter

\newcounter{author}
\renewcommand{\author}[2][]{
   \stepcounter{author}
   \@namedef{author@\theauthor}{#2}
   \@namedef{authorlabel@\theauthor}{#1}
}

\newcounter{address}
\newcommand{\address}[2][]{
   \stepcounter{address}
   \@namedef{address@\theaddress}{#2}
   \@namedef{addresslabel@\theaddress}{#1}
}

\newcommand{\alsep}{and}

\def\newmaketitle{\par%
  \begingroup%
  \normalfont%
  \def\thefootnote{}
  \def\footnotemark{}
  \let\@makefnmark\relax
  \footnotesize
  \footnotesep 0.7\baselineskip
  \normalsize%
  \twocolumn[\thenewmaketitle\@IEEEaftertitletext]%
  \if@IEEEusingpubid
     \enlargethispage{-\@IEEEpubidpullup}%
  \fi
  \endgroup
  \setcounter{footnote}{0}\let\maketitle\relax\let\@maketitle\relax
  \gdef\@thanks{}%
  \let\thanks\relax}

\def\thenewmaketitle{
  \newpage
  \begin{center}%
    \vskip0.2em{\Huge\@IEEEcompsoconly{\sffamily}\@IEEEcompsocconfonly{\normalfont\normalsize\vskip 2\@IEEEnormalsizeunitybaselineskip
   \bfseries\large}\@title\par}\vskip1.0em\par%
    \vspace{1ex}
    \newcounter{c@author}
    \newcounter{c@tmp}
    \ifthenelse{\value{author}=2}{%
      \newcommand{\liand}{ and }}{%
      \newcommand{\liand}{, and }}
    \ifthenelse{\value{address}<2}{%
      \@nameuse{author@1}%
      \stepcounter{c@author}%
      \whiledo{\value{c@author}<\value{author}}{%
        \setcounter{c@tmp}{\value{author}}%
        \addtocounter{c@tmp}{-\value{c@author}}%
        \ifthenelse{\value{c@tmp}=1}{%
          \renewcommand{\alsep}{\liand}}{\renewcommand{\alsep}{, }}%
        \stepcounter{c@author}\alsep \@nameuse{author@\thec@author}}\\%
    }
    {
      \@nameuse{author@1}${}^{(\ref{\@nameuse{authorlabel@1}})}$%
      \stepcounter{c@author}%
      \whiledo{\value{c@author}<\value{author}}{%
      \setcounter{c@tmp}{\value{author}}%
      \addtocounter{c@tmp}{-\value{c@author}}%
      \ifthenelse{\value{c@tmp}=1}{%
        \renewcommand{\alsep}{\liand}}{\renewcommand{\alsep}{, }}%
      \stepcounter{c@author}\alsep \@nameuse{author@\thec@author}%
        ${}^{(\ref{\@nameuse{authorlabel@\thec@author}})}$%
      }
    }
    \vspace{0.2ex}

    \ifthenelse{\value{address}>0}{%
      \ifthenelse{\value{address}=1}{
        {\@nameuse{address@1}}
      }
      {
        \newcounter{c@address}

        \begin{center}
        \whiledo{\value{c@address}<\value{address}}
        {
          \refstepcounter{c@address}
            ${}^{(\thec@address)}$\,%
              \label{\@nameuse{addresslabel@\thec@address}}%
              \@nameuse{address@\thec@address}\\ %
        }
        \end{center}
      } 
    }
    {
      \relax
    }
  \end{center}
}

\makeatother

\title{An Electrodynamics Solver for Moving Sources}

\author[org1]{Sameh Y.~Elnaggar}
\author[org1]{Yahia M.~M.~Antar}

\address[org1]{The Royal Military College of Canada,\\ Email: S.Y.E (samehelnaggar@ieee.org), Y.~M.~M.~A (antar-y@rmc.ca)}

\begin{document}

\newmaketitle

\begin{abstract}
An Electrodynamics solver for moving sources is introduced. The main challenges and formulation are highlighted. The solver enables the simulation of fields for sources undergoing arbitrary motion. Two examples of uniformly moving current sources are provided to correlate the numerical solver computations with theory, based on the solution of Maxwell's equations and the relativistic transformation of the electromagnetic fields.
\end{abstract}

\section{Introduction}

Computational Electromagnetic (CEM) tools have been growing over the last three decades. They are routinely used by engineers and scientists to simulate complex systems such as antenna arrays, radars, multi-layer printed circuit boards, and plasmas. Free available and commerical tools provide platforms that enable the rapid setting of such complex systems \cite{oskooi2010meep,davidson2010computational,sumithra2017review,taflove2013advances}. From a pedagogical perspective, instructors leverage CEM tools to make the theory of electromagnetism more tangible and accessible to students. Although some of these tools take into account the effect of moving objects, their scope is usually limited to specific cases or certain operating regimes. For instance magneto-statics is usually coupled with Lorentz force to enable the simulation of motors in the sub-wavelength regime. Additionally, the full Maxwell's equations are coupled with Newton's second law to study the behaviour of plasma gaseous, where the charged particles are usually considered to be point-wise \cite{arber2015contemporary}. 

In the current article, we provide a general overview of a CEM solver that exploits the global form interpretation of Maxwell's equations to intuitively and rigorously compute the fields of an aribitrary number of moving sources. Sources are considered to be macroscopic harmonic current distributions of the form $\sum\mathbf{J}_k(\mathbf{r})\exp(-i\omega_k t)$. So far, we only consider non-relativistic sources, i.e, current distributions are not functions of the moving speed $\mathbf{v}$.

In seciton II, the challenges and general theme of the solver are highlighted. The global form of Maxwell's equations in freespace is presented. The form readily suggests a convienent framework to describe  the continuous motion of sources. Additionally, the steady state and transient solvers are introduced. It will be shown that the transition from the stationary to the transient fields necessitates the generalization of the Perfectly Matching Layer (PML) formalism. Section III gives two canonical examples of 2D uniformly moving sources. The numerical calculations are compared to an exact analytical model, where a closed form of the fields are derived in the proper frame of reference and then transformed using Lorentz transformation to the observed frame.
\section{Challenges and Solver main components}
The solver leverages many of the already developed CEM techniques. However to taken into account the continuous movement of current sources, adaptations and extensions are inevitable. The following two subsections present a brief description of two main challenges and how they are handled.

\subsection{Capturing continuous motion}
The reported solver exploits the well developed machinery of the finite difference frequence domain (FDFD)  \cite{rumpf2022electromagnetic} and finite difference time domain (FDTD) \cite{taflove2005}. Nevertheless to correctly model the continuous motion of sources, some modifcations and re-interpretation of the different terms in the FDTD update equations are necessary. 

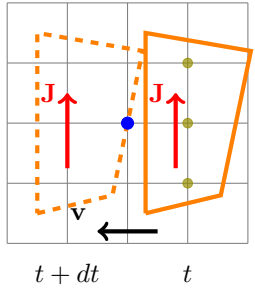
\begin{figure}
\centering
\begin{tikzpicture}[scale=0.8]
\foreach \x in{-2,-1,0,1,2}
{
	\draw [gray,thin](\x,-2)--(\x,2);
}
\foreach \y in{-2,-1,0,1,2}
{
	\draw [gray, thin](-2,\y)--(2,\y);
}
\draw [orange,ultra thick, dashed](-1.5,-1.5)--(-0.25,-1.2)--(0.25,1.2)--(-1.5,1.5)--(-1.5,-1.5);
\def \sh{1.8};
\draw [orange,ultra thick](-1.5+\sh,-1.5)--(-0.25+\sh,-1.2)--(0.25+\sh,1.2)--(-1.5+\sh,1.5)--(-1.5+\sh,-1.5);
\draw [ultra thick,->](0.5,-1.8)--(-0.5,-1.8) node[anchor=south east]{ $\mathbf{v}$};
\node at (1,-2.5){$t$};
\node at (-1,-2.5){$t+dt$};
\draw[blue, fill=blue](0,0) circle(0.1);
\draw [ultra thick, red,->](-1,-0.75)--(-1,0.5) node[anchor=east]{$\mathbf{J}$};
\draw [ultra thick, red,->](-1+\sh,-0.75)--(-1+\sh,0.5) node[anchor=east]{$\mathbf{J}$};
\draw [olive, fill=olive,opacity=0.7](1,0) circle(0.08);
\draw [olive, fill=olive,opacity=0.7](1,1) circle(0.08);
\draw [olive, fill=olive,opacity=0.7](1,-1) circle(0.08);
\end{tikzpicture}
\caption{The position of a current source in the plane at two instants $t$ and $t+dt$. The dot represents a grid point, where in current is defined in a typical FDTD formalism.}
\label{fig::fig1}
\vspace{-5mm}
\end{figure}

Figure \ref{fig::fig1} shows a current sheet that is moving with a velocity $\mathbf{v}$ in the plane. From an FDFD (or similarly FDTD) point of view, the current is defined at the grid points. Hence a finite difference paradigm represents a source by its discretized samples in space as the figure highlights. To accurately represent the current distribution at a given instant, the spatial sampling frequency must be high enough (at least double the highest distribution spatial frequency). For distributions with a large gradient a dense grid becomes essential. Furthermore when the source moves, the sampling points will change. Consider for example, the situation in Fig. \ref{fig::fig1} at $t+dt$. At this instant, the highlighted blue point \emph{sees} a current value. Ones the object slightly moves to the left, the current at the same point drops to zero. This switching on and off effect results in a jittered and noisy performance and is equivalent to non-physical creation and annhilation of charges. 
\begin{table*}
\centering
\caption{\fontsize{9}{11}\selectfont{Summary of Main Equations of Electromagnetism in the time domain. 
$\rho$ charge density, $\mathbf{J}$ current density, $Q_f$ total charge flow in $t_2-t_1$, $Q(t_i)$ charge inside volume at $t=t_i$, $\mathbf{D}$ electric flux density, $\Phi$ electric flux across surface, $\mathbf{H}$ magnetic field intensity, $\langle\textnormal{mmf}\rangle$ average magnetomotive force from $t_1$ to $t_2$, $\mathbf{E}$ electric field intensity, $\mathbf{B}$ magnetic flux density, $\langle\textnormal{emf}\rangle$ average electromotive force from $t_1$ to $t_2$, and $\Psi$ magnetic flux across surface.}}
\label{table::table1}
\rotatebox{00}{
\begin{tabular}{|c|c|c|c|}
\hline
Law & Differential & Integral & Global \\
\hline\hline
Cont. of Q & $\nabla\cdot\mathbf{J}=-\frac{\partial \rho}{\partial t}$ & $\iint_{S} \mathbf{J}\cdot d\mathbf{S}=-\frac{\partial{Q}}{\partial t}$ & $Q_f=Q(t_1)-Q(t_2)$ \\
\hline
Gauss E  &$\nabla\cdot\mathbf{D}=\rho $& $\iint_S\mathbf{D}\cdot d\mathbf{S}=Q$ & $\sum_\textnormal{all faces}\Phi=Q$ \\
\hline
Ampere's & $\nabla\times \mathbf{H}=\mathbf{J}+\frac{\partial\mathbf{D}}{\partial t}$ & $\oint\mathbf{H}\cdot d\mathbf{l}=\iint_S \mathbf{J}\cdot d\mathbf{S}+\iint_S\frac{\partial \mathbf{D}}{\partial t}\cdot d\mathbf{S}$ & $\langle\textnormal{mmf}\rangle=Q_f+\frac{\Phi(t_2)-\Phi(t_1)}{t_2-t_1}$ \\
\hline
Faraday's & $\nabla\times \mathbf{E}=-\frac{\partial B}{\partial t}$ & $\oint \mathbf{E}\cdot d\mathbf{l}=-\iint_S\frac{\partial \mathbf{B}}{\partial t}\cdot d\mathbf{S}$ & $\langle\textnormal{emf}\rangle=-\frac{\Psi(t_2)-\Psi(t_1)}{t_2-t_1}$ \\
\hline
Gauss B & $\nabla\cdot\mathbf{B}=0$ & $\iint_S\mathbf{B}\cdot d\mathbf{S}=0$ & $\sum_\textnormal{all faces}\Psi=0$\\
\hline
\end{tabular}
}
\end{table*}
To capture the continuous movement of sources without excessively increasing the grid size, we take a step back and consider the three equivalent forms of Maxwell's equations shown in table \ref{table::table1}. The differential and integral forms are widely used. The third form is derived from the integral representation by integrating over a time interval $t_2-t_1$ and representing flux quantities by their \emph{average} flux density counterparts. Such form is not new of course and has been widely used in CEM formalisms \cite{tonti2013mathematical,clemens2001discrete}. Since the form relates average quantities to one another, it is also very convenient when correlating results to measurements. Measured quantities are macroscopic by nature and represent average values\cite{tonti2013mathematical}. Using Algebraic Topology, it was shown that most of the physical theories including electromagnetism can be formulated in a discrete form that associate variables with points, lines, surfaces and volumes \cite{tonti2013mathematical}.

The global formalism suggests the use of the total charge flow $Q_f$ through the Yee cell surfaces during the interval $t_2-t_1$ instead of the discrete sampled current at the grid points and time instant $(t_1+t_2)/2$. As a source moves through the grid, its intersection with the different surfaces (or edges in 2D) is continuous; hence charge flow is continuously captured.

The above discussion implies the necessity of calculating the intersection area between a moving source and  a grid surface. Such process can be quite involved for complex sources geometries in 3D. In our solver, we use an open source efficient computational geometry library (\textbf{CGAL}) \cite{fogel2015computational} to perform such geometrical operations. CGAL is a template C++ library that provides thousands of compuational geometry predicates and subroutines. It also enables the use of exact numerical types that are essential for the reliable operation of some critical predicates (for example, point inclusion predicates).

\subsection{Steady State and Transient Solvers}
We consider sources to be stationary at $t<0$. They may move in the interval $t\geq 0$. Hence for $t<0$, the stationary form of Maxwell's equations as represented by the global formalism is used (table \ref{table::table2}), which is derived from the global form in table \ref{table::table1} after replacing the time dependency by $\exp(-i\omega t)$. Unlike the time dependent form, the stationary form is a system of algebraic equations that is formulated using the Finite Difference Frequency Domain (FDFD) technique as
\begin{equation}
\label{eq:ABCD}
\begin{bmatrix}
\mathbf{A}(\omega)  & \mathbf{B}(\omega)\\
\mathbf{C}(\omega) &  \mathbf{D}(\omega)
\end{bmatrix}
\begin{bmatrix}
\mathbf{E}(\omega) \\
\mathbf{H}(\omega)
\end{bmatrix}
=
\begin{bmatrix}
\mathbf{I}(\omega)\\
\mathbf{0}
\end{bmatrix}.
\end{equation}
The $\mathbf{A}$, $\mathbf{B}$, $\mathbf{C}$ and $\mathbf{D}$  matrices are sparse. Eigen, an open source linear algebra, is exploited to efficiently calculate the system of equations (\ref{eq:ABCD})\cite{eigenweb}.

To model the behaviour of unbounded space, where travelling waves do not bounce back from the inevitable truncated computational space, the stretched theory formalism of perfectly matched layers is exploited \cite{chew19943d,taflove2005computational}. The spatial dimensions are scaled by complex single pole transfer functions to assure that waves attenuate as they move into the layers. In the time domain, the interaction between the waves and the layers is presented by a convolutional operator. Traditionally, initial fields values are assumed to be zero. In our case, however, fields do exist and fill all space for $t<0$. Therefore, the formalism is modified to include the effect of interaction of the fields at $t\leq 0$ with the convolution operator. The effect of the steady state fields appears as an extra term in the update equations, which exponentially goes to zero over time and is recursively calculated. This means that no extra run time over head is incurred in the update loop.

The system of update equations is formulated in a matrix form that can be written as\vspace{-1mm}
\begin{equation}
\mathbf{X}_n=\mathbf{X}_{n-1}+\mathbf{\mathcal{L}}\mathbf{Y}_n+\mathbf{\mathcal{I}}_n,
\end{equation}
where $\mathbf{X}_{n}$ and $\mathbf{Y}_n$ are the $n^\textnormal{th}$ updated fields $\mathbf{X}$ and $\mathbf{Y}$ that are either $\mathbf{E}$ or $\mathbf{H}$. The matrix $\mathbf{\mathcal{L}}$ represents the $\nabla\times$ operator, which is sparse. Additionally, the vector $\mathbf{\mathcal{I}}$ represents the source term, appearing in Ampere's law. The matrix form delegates the update computations to the linear algebra subroutines that may exploit the vectorization capabilities of the underlying hardware architecture. 
\begin{table}[h]
\centering
\caption{\fontsize{9}{11}\selectfont{Global form for harmonic sources, $\tilde{m}$ ($m_0$) is the amplitude of the magnetomotive force at frequency $\omega$ (DC). Similarly, $\tilde{e}$ and $e_0$ are amplitude of the electromotive force at $\omega$ and the electrostatic emf, respectively.}}
\label{table::table2}
\begin{tabular}{|c||c|}
\hline
Law & Form\\
\hline\hline
Continuity &$\sum_{\partial V} \tilde{I}=i\omega\tilde{Q}$\\
\hline
Electric Gauss  &$\sum_{\partial V}\tilde{\Phi}=\tilde{Q},~~\sum_{\partial V}{\Phi}_0={Q}_0$\\
\hline
Magnetic Gauss &$\sum_{\partial V}\tilde{\Psi}=0$\\
\hline
Maxwell-Ampere & $\sum_\circlearrowleft\tilde{m}=\tilde{I}-i\omega\tilde{\Phi}$, $\sum_\circlearrowleft m_0=I_0$\\
\hline
Maxwell-Faraday & $\sum_\circlearrowleft\tilde{e}=i\omega\tilde{\Psi}$, $\sum_\circlearrowleft e_0=0$\\
\hline

\end{tabular}
\end{table}

\section{Numerical Examples}
\vspace{-1mm}
In this section, we consider two examples of uniformly moving sources in the $x$ direction. The sources are homogeneous in the $z$ direction ($\partial_z=0$); hence reducing Maxwell's equations to TE and TM independent sets of equations. The first example presents the results of a singular delta source, where current is flowing out of the plane. In the second example, we treat the case of a thin circular ring carrying a uniform current. Throughout this section, Maxwell's equations are re-written in normalized units such that $\epsilon_0=1$, $\mu_0=1$ and hence the speed of light $c=1$. Additionally, time instants $t$ are represented in terms of the equivalent lengths $ct$. This means that $t=1$ m means that $t$ is equivalent to the time in seconds a light ray needs to travel a distance of 1 m.

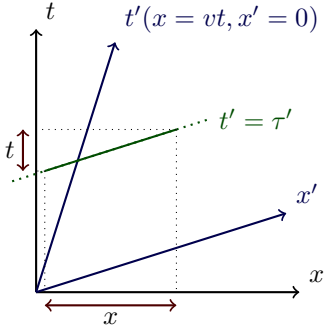
\begin{figure}
\centering
\begin{tikzpicture}[scale=1.75]
\draw [thick,->] (0, 0)--(2,0) node [anchor=south west]{$x$};
\draw [thick,->] (0, 0)--(0,2) node [anchor=south west]{$t$};
\draw [thick,->,darkblue](0,0)--(1.9,0.6) node [anchor=south west]{$x'$};
\draw [thick,->,darkblue](0,0)--(0.6,1.9) node [anchor=south west]{$t' (x=vt, x'=0)$};
\draw [thick,darkgreen,dotted](1*0.6/1.9,1)--+(1,0.6*1/1.9) node[anchor=west]{$t'=\tau'$};
\draw [thick,darkgreen,dotted](1*0.6/1.9,1)--+(-0.5,-0.6*0.5/1.9);
\draw [thick,darkgreen](1*0.6/1.9,1)--+(0.75,0.6*0.75/1.9);
\draw [thick,darkgreen](1*0.6/1.9,1)--+(-0.25,-0.6*0.25/1.9);
\draw [thin,dotted](1*0.6/1.9+0.75,1+0.6*0.75/1.9)--(1*0.6/1.9+0.75,0);
\draw [thin, dotted](1*0.6/1.9-0.25,1-0.6*0.25/1.9)--(1*0.6/1.9-0.25,0);
\draw [darkred,<->, thick](1*0.6/1.9+0.75,-0.1)--(1*0.6/1.9-0.25,-0.1);
\node at (1*0.6/1.9+0.25,-0.2){$x$};
\draw [thin,dotted](1*0.6/1.9+0.75,1+0.6*0.75/1.9)--(0,1+0.6*0.75/1.9);
\draw [thin, dotted](1*0.6/1.9-0.25,1-0.6*0.25/1.9)--(0,1-0.6*0.25/1.9);
\draw [thick,darkred,<->](-0.1,1-0.6*0.25/1.9)--(-0.1,1+0.6*0.75/1.9);
\node at(-0.2,1+0.6*0.25/1.9){$t$};
\end{tikzpicture}
\caption{Minkowski Diagram highlighting the grid ($S$) and proper ($S'$) reference frames. The green line represents the $x'$ coordinate as observed in $S'$ for proper time $t'=\tau'$. The projections on the $S$ $x$ and $t$ coordinates of the space-time events represented by the solid green line. }
\label{fig:spacetime}
\vspace{-3mm}
\end{figure}

In both examples, Maxwell's equations are solved in the proper frame of reference ($S'$), where the sources are stationary. Lorentz transformation is then applied to calculate the fields in the grid frame ($S$). The transformation is performed in two stages. In the first stage, the events $(x,y,t)$ that represent the computational domain as observed in $S$ are transformed to the corresponding ones $(x',y',t')$ in the $S'$ frame. Figure \ref{fig:spacetime} shows Minkowski's diagram that gives a pictorial description of Lorentz transformation. It is worth noting that the simultaneous events $(x',t'=\tau')$, representing the wave-front at $t'=\tau'$ are not simultaneous in $S$; a direct consquence of the relative simulatenity principle in Special relativity. Such behaviour appears as a change in the observed frequency (Doppler effect).  
Explicitly, the space-time coordinates in $S'$ and $S$ are related by the well-known relations:\vspace{-1mm}
\begin{align*}
x'&=\gamma(x-vt),~
y'=y,~
t'=\gamma(t-vx),~
\gamma \triangleq(1-v^2)^{-1/2}.
\end{align*}
The second step is to transform the fields observed in $S'$ to the ones in $S$ using the transformation of fields \cite{griffiths2005}
\begin{align}
\mathbf{E_\perp}&=\gamma\left(\mathbf{E_\perp'}-\mathbf{v}\times\mathbf{B_\perp'}\right),~~
\mathbf{B_\perp}=\gamma\left(\mathbf{B_\perp'}+\mathbf{v}\times\mathbf{E_\perp'}\right)
\end{align}

\subsection{\vspace{-1mm}Out of plane singular source}
From Maxwell's equations, it can be shown that the fields of a singular harmonic located at the origin with an angular frequency $\omega$ can be written as
\begin{align}
E_z'&=-\frac{\omega}{4}I_0\mathcal{H}_0(\omega r')e^{-i\omega t}+c.c,\\
H_\phi'&=-\frac{i}{4}I_0\mathcal{H}_1(\omega r')e^{-i\omega t}+c.c,
\end{align}
where $I_0$ is the total current and $r'\triangleq \sqrt{{x'}^2+{y'}^2}$ is the distance from the origin as observed in $S'$.
\begin{figure}
\centering
\begin{tikzpicture}[scale=2]
\foreach \x in{-0.8,-0.7,...,0.8}
{
\draw [gray, thin](-0.8,\x)--(0.8,\x);
\draw [gray,thin](\x,-0.8)--(\x,0.8);
\draw [blue,opacity=0.1,fill=blue](0,0) circle(0.2);
\draw [blue,opacity=0.1,dotted,fill=blue](0.5,0) circle(0.2);
\draw [orange, fill=orange,opacity=0.05](-0.8,0.8)rectangle+(1.6,0.25);
\draw [orange, fill=orange,opacity=0.05](0.8,-0.8)rectangle+(0.25,1.85);
\draw [orange, fill=orange, opacity=0.05](-0.8,-0.8) rectangle+(1.85,-0.25);
\draw [orange, fill=orange, opacity=0.05](-0.8,-1.05)rectangle+(-0.25,2.1);
\node at (0,0.9){\scriptsize{PML}};
\draw [ultra thick,->](0.2,-0.3)--(0.4,-0.3)node[anchor=north]{$v$};
\draw [black,opacity=0.1,fill=black](0,0) circle(0.04);
\draw [black,opacity=0.1,dotted,fill=black](0.5,0) circle(0.04);
\node [white]at(0.5,0.09){$J_z$};
\node [white]at(0,0.09){$J_z$};
}
\end{tikzpicture}
\caption{Computational domain of an out of plane current.}
\label{fig:TM}
\vspace{-3mm}
\end{figure}
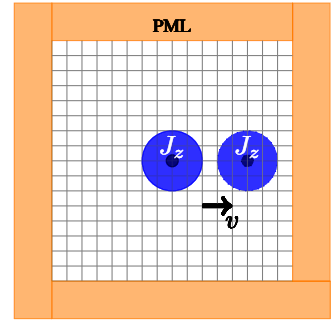
Figure \ref{fig:TM} shows the computational domain used in the calculation. The radius of the current source $R$ is much smaller than the wavelength to approximate the singular source. However, it must be larger than the grid size to continuously capture the effect of movement on the fields. Additionally, the current density $J_z$ is chosen such that $J_z\pi R^2=I_0$. We have considered the source to be rigid and hence neglecting Lorentz contraction. This implies that the solver operates in the non-relativistic regime, where $v\ll 1$. Nevertheless to highlight the effect of motion on the fields, we let $v=0.4$. The source is stationary for $t<0$. At this instant, as Fig. \ref{fig:TMfields}(a) shows the fields are circularly symmetric. Nevertheless as the source moves, the fields change shape as shown in Fig. \ref{fig:TMfields}(b), where the observed frequency increases to the right of the source and expands to the left as a reflection of Doppler effect.
\begin{figure}
\centering
\begin{subfigure}{0.47\linewidth}
\includegraphics[width=\linewidth]{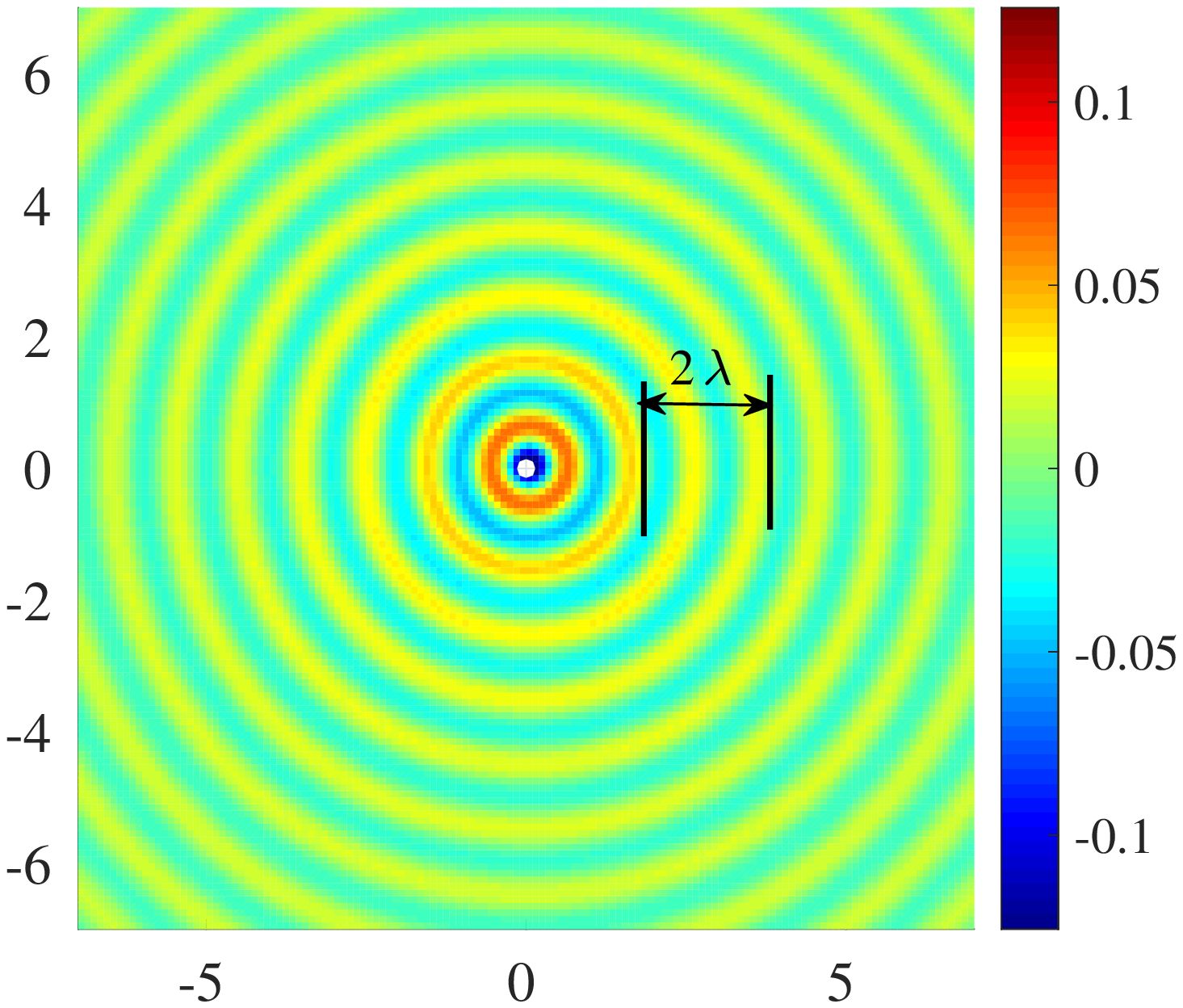}
\caption{$t=0$}
\end{subfigure}
\begin{subfigure}{0.47\linewidth}
\includegraphics[width=\linewidth]{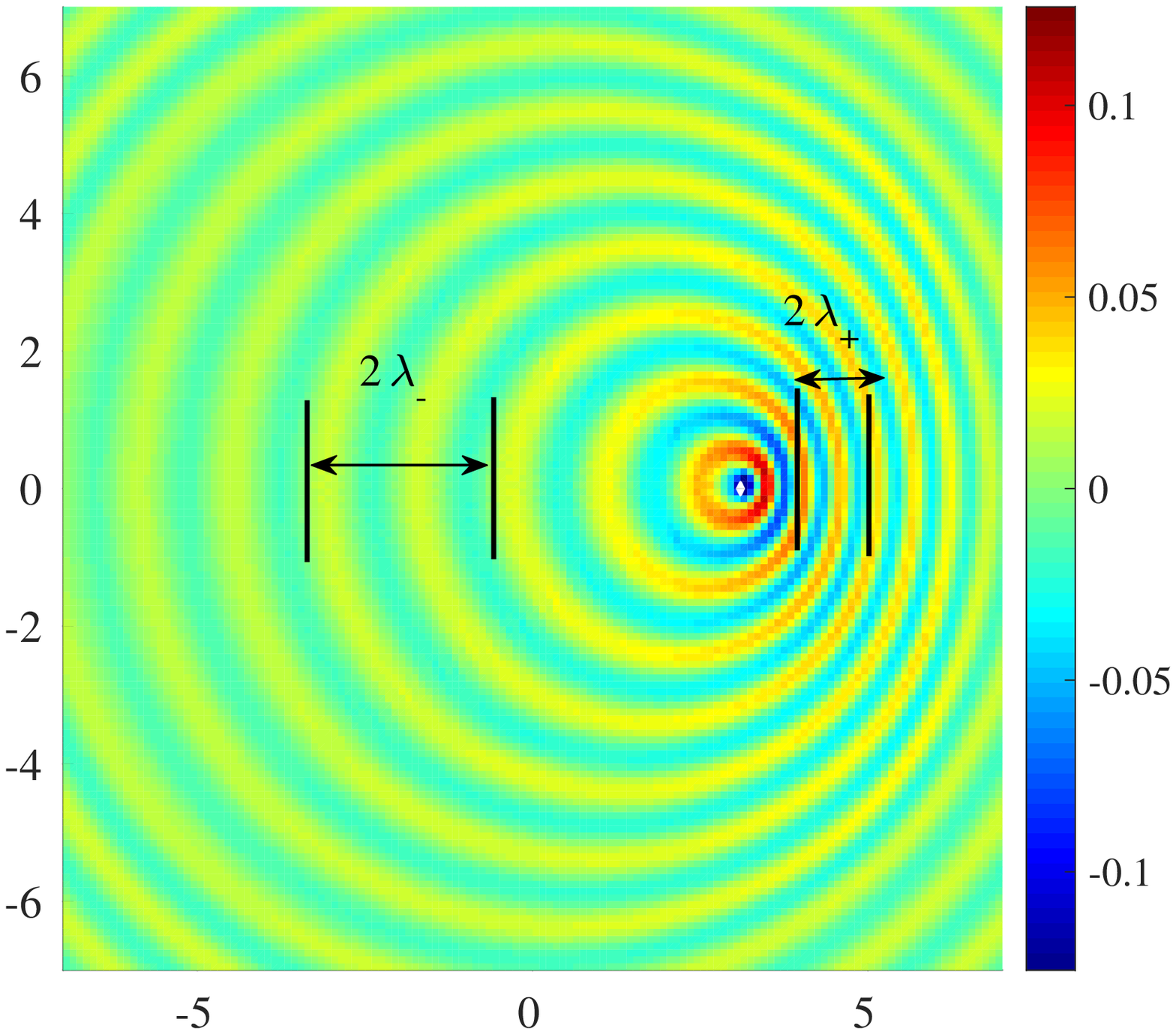}
\caption{$t=7.9$ m}
\end{subfigure}
\caption{$E_z$ as observed in $S$ at two instants. (a) At $t=0$ when the source starts to move and (b) at $7.9$ m normalized time units after the source moved.}
\label{fig:TMfields}
\vspace{-2mm}
\end{figure}
\subsection{2D circular ring}
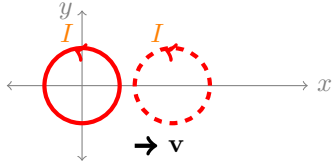
\begin{figure}
\centering
\begin{tikzpicture}
\draw [gray,thin,<->](-1,0)--(3,0)node[anchor=west]{$x$};
\draw [gray, thin,<->](0,-1)--(0,1)node[anchor=east]{$y$};
\draw [red,ultra thick](0,0)circle(0.5);
\draw [red,ultra thick, dashed](1.2,0)circle(0.5);
\draw [red,ultra thick,->](0,0.5)to[bend left](-0.1,0.48);
\draw [red,ultra thick,->,dashed](1.2,0.5)to[bend left](1.1,0.48);
\node [orange]at (-0.2,0.7){$I$};
\node [orange]at(1,0.7){$I$};
\draw [ultra thick,->](0.7,-0.8)--(1,-0.8)node[anchor=west]{$\mathbf{v}$};
\end{tikzpicture}
\caption{A unformly moving thin 2D circular ring.}
\label{fig::fig2}
\vspace{-3mm}
\end{figure}
We consider the 2D thin ring shown in Fig. \ref{fig::fig2}, which carries a uniform current $I$. In an inertial frame of reference co-moving with the loop, one can find expressions for the fields by solving Maxwell's equations where $\partial_z=0$ and after imposing the boundary conditions at $r=r_0$ to show that
\begin{equation}
H_z'(r,t)=-ie^{-i\omega t'}\begin{cases}
AJ_0(\omega r') ,~r'\leq r_0\\
C\mathcal{H}_0(\omega r'),~r'>r_0
\end{cases}
\end{equation}
and\vspace{-2mm}
\begin{equation}
E_\phi'(r,t)=e^{-i\omega t'}\begin{cases}
AJ_1(\omega r') ,~r'\leq r_0\\
C\mathcal{H}_1(\omega r'),~r'>r_0
\end{cases},
\end{equation}
where $J_n(x)$ is the $n^\textnormal{th}$ Bessel function, $\mathcal{H}_n(x)$ is Hankel function of first kind. The constants $C$ and $A$ are related via the continuity of $E_\phi$ at $r=r_0$, i.e,
\begin{equation}
C=\frac{J_1(\omega r_0)}{\mathcal{H}_1(\omega r_0)}A.
\end{equation}
The constant $A$ is a function of the total current $I$, i.e,
\begin{equation}
A=\frac{i\mathcal{H}_1(\omega r_0)}{ J_0(\omega r_0)\mathcal{H}_1(\omega r_0)-J_1(\omega r_0)\mathcal{H}_0(\omega r_0) }I.
\end{equation}
In the computational frame, the current loop moves with a uniform velocity $\mathbf{v}$ in the $x$ direction. Therefore, the fields in $S$ can be computed as has been described earlier in this section.

Since the solver does not take into account the relativisitc effects such as Lorentz contraction, we expect inaccurate results. Therefore for a quantitative analysis, we limit $v$ to be less than 0.01 (equivalently $v=0.01c$).

Furthermore we consider a loop of normalized radius $r_0=0.8$ and a normalized frequency $f=1$. The loop thickness is $0.1$ units, which must be larger than the grid cells linear length (chosen to be $0.05$ units). The loop starts to move at $t=0$ in the positive $x$ direction. 

\begin{figure}
\begin{subfigure}[t]{0.22\textwidth}
\includegraphics[width=\textwidth]{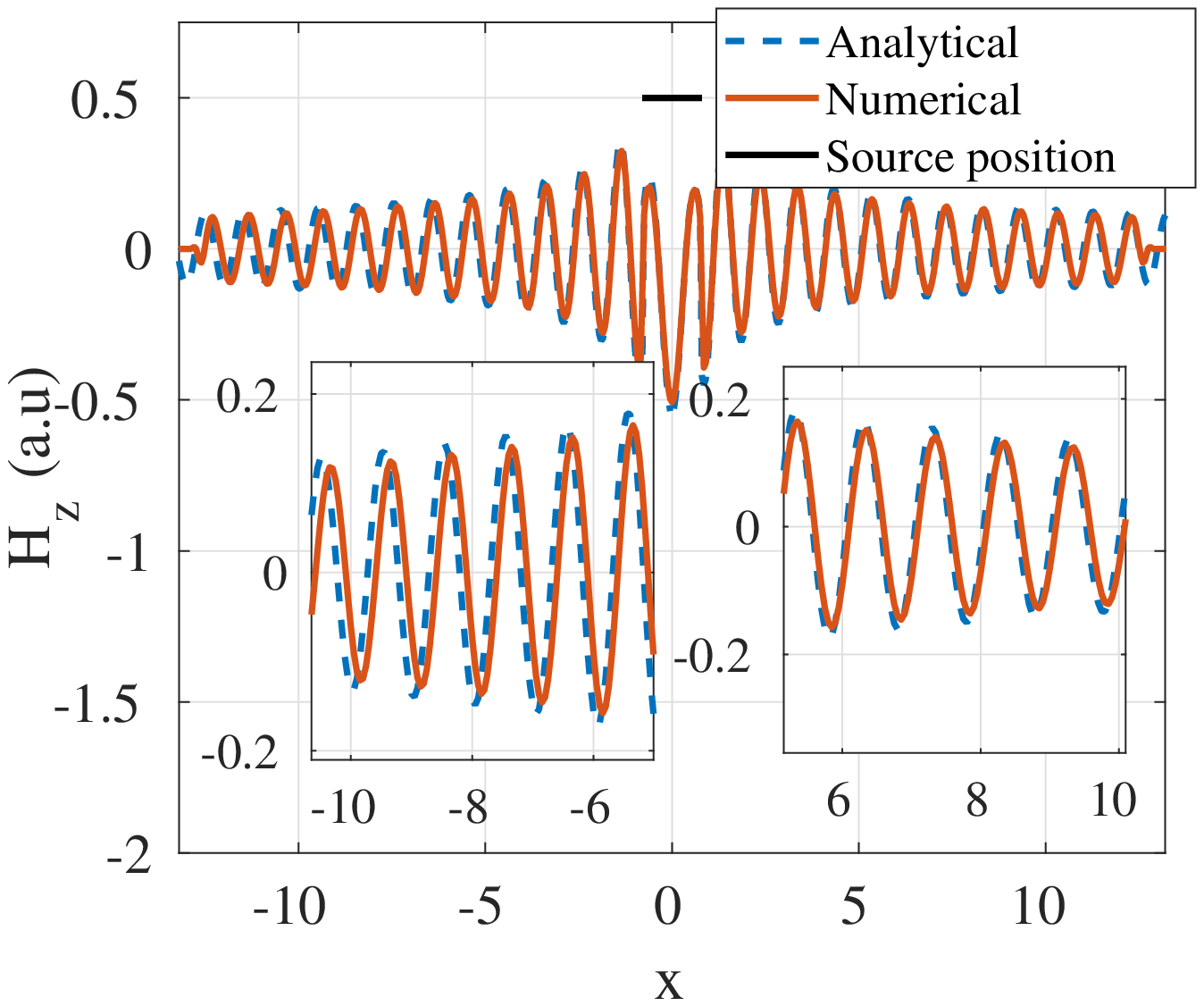}
\caption{$t=0$}
\end{subfigure}
\begin{subfigure}[t]{0.22\textwidth}
\includegraphics[width=\textwidth]{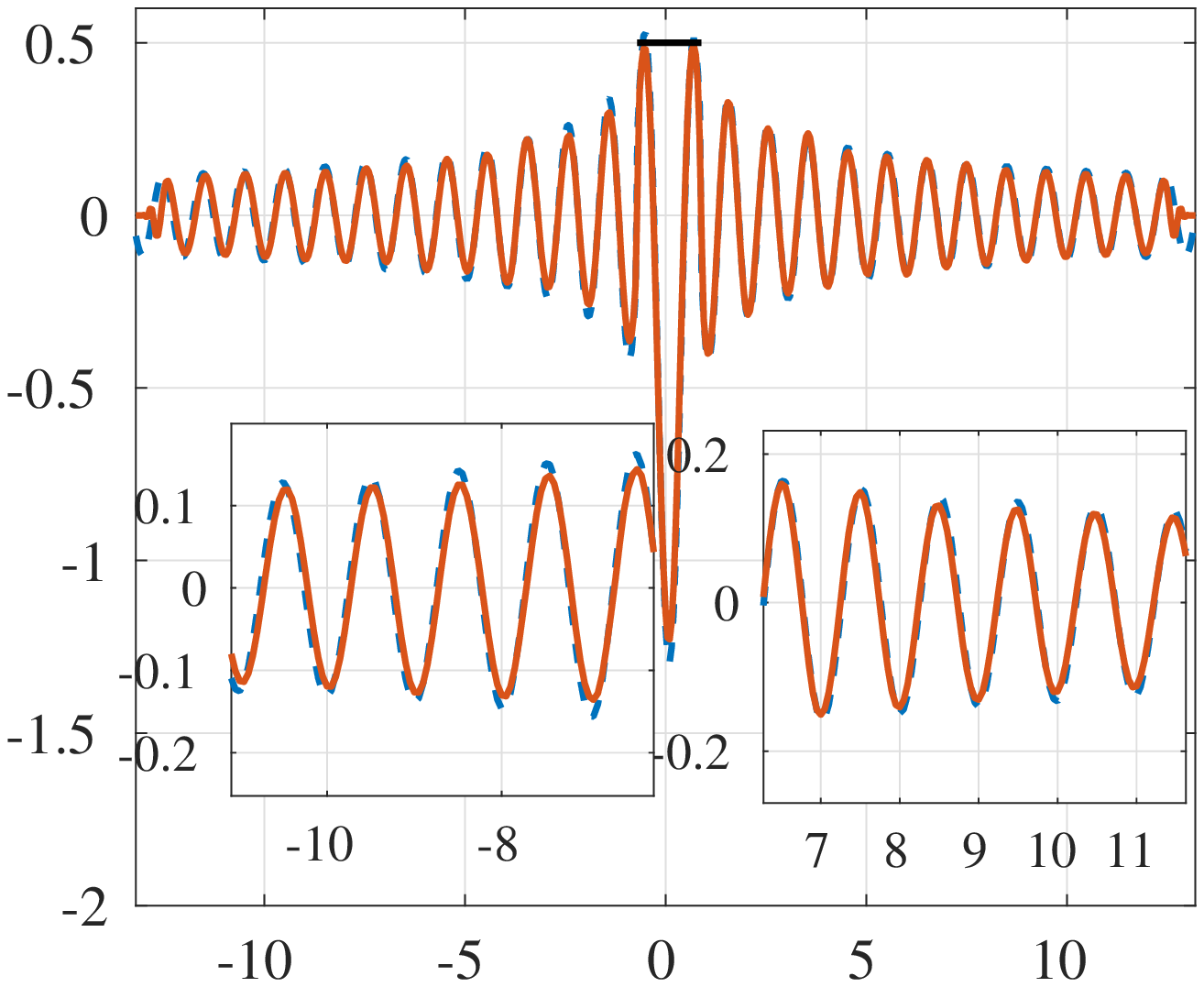}
\caption{$t=9.12 \textnormal{ m}$}
\end{subfigure}
\caption{$H_z$ at two different time instants.}
\vspace{-3mm}
\label{fig:Hz_x}
\end{figure}

Figure \ref{fig:Hz_x} presents the computed $H_z$ on the $x$ axis and compared with the analytical calculations. Initially when the source is about to move, there is a phase shift between the analytical model and the numerical computations. Eventually, the wavelength in the grid frame changes due to Doppler effect and the computed and analytical phases become synchronized as Fig. \ref{fig:Hz_x}(b) demonstrates.
\section{Conclusion}
We introduced a general description of an electrodynamics solver to simulate moving harmonic sources. Using the global formalism and computational geometry algorithms, the solver enables the simulation of the fields for sources undergoing arbitrary motion. It was applied to two canonical examples that have closed form expressions.   Besides seeking answers to fundamental questions about reaction force and energy in near the near field, the solver is expected to be a vital tool in the analysis of moving antennas appearing in tomorrow's technologies such as internet of things and 5+ G networks.
\bibliographystyle{IEEEtran}
\bibliography{CEM}
\end{document}